\documentclass[aps,prd,twocolumn,superscriptaddress,amsmath,amssymb]{revtex4}
\usepackage{graphicx,overpic} % Include figure files
\usepackage{epstopdf} % Allow to include eps files
\usepackage{dcolumn} % Align table columns on decimal point
\usepackage{xcolor} % Color support
\usepackage{hyperref} % Generate pdfs with hyperlinks
\usepackage[utf8]{inputenc} % Allow UTF8 characters
\usepackage[english]{babel} % All publications are in English
\usepackage[mathlines]{lineno} % Line numbers for drafts
\usepackage{blindtext} % For testing
\usepackage{amsmath} % Adds a large collection of math symbols
\usepackage{amssymb}
\usepackage{amsfonts}
\usepackage{upgreek} % Adds in support for greek letters in roman typeset
\usepackage[normalem]{ulem}

\usepackage{orcidlink}

\newcommand*\patchAmsMathEnvironmentForLineno[1]{%
\expandafter\let\csname old#1\expandafter\endcsname\csname #1\endcsname
\expandafter\let\csname oldend#1\expandafter\endcsname\csname
end#1\endcsname
 \renewenvironment{#1}%
   {\linenomath\csname old#1\endcsname}%
   {\csname oldend#1\endcsname\endlinenomath}%
}
\newcommand*\patchBothAmsMathEnvironmentsForLineno[1]{%
  \patchAmsMathEnvironmentForLineno{#1}%
  \patchAmsMathEnvironmentForLineno{#1*}%
}
\AtBeginDocument{%
\patchBothAmsMathEnvironmentsForLineno{equation}%
\patchBothAmsMathEnvironmentsForLineno{align}%
\patchBothAmsMathEnvironmentsForLineno{flalign}%
\patchBothAmsMathEnvironmentsForLineno{alignat}%
\patchBothAmsMathEnvironmentsForLineno{gather}%
\patchBothAmsMathEnvironmentsForLineno{multline}%
\patchBothAmsMathEnvironmentsForLineno{eqnarray}%
}

\graphicspath{{figures/}} % Use figures directory for figures

\usepackage{relsize}
\def\babar{\mbox{\slshape B\kern-0.1em{\smaller A}\kern-0.1em
    B\kern-0.1em{\smaller A\kern-0.2em R}}\xspace}

\newcommand{\lumi}{\ensuremath{207\invfb}\xspace}

\newcommand{\OmgcToOmgPi}{\ensuremath{\Omgc\to\Omg\pip}\xspace}

\newcommand{\mOmgc}{\ensuremath{m(\Omg\pip)}\xspace}

\newcommand{\sigmat}{\ensuremath{\sigma_t}\xspace}

\newcommand{\pcms}{\ensuremath{p_{\rm cms}}\xspace}

\newcommand{\tauOmgc}{243}
\newcommand{\tauOmgcStat}{48}
\newcommand{\tauOmgcSyst}{11}

\input{belle2sym}

\usepackage{ifthen} % for conditional statements

\newboolean{wordcount}
\setboolean{wordcount}{false} % False for normal usage; true for wordcount.sh

\ifthenelse{\boolean{wordcount}}%
{\usepackage{bibentry} % For nobibliography command
 \usepackage{comment} % comment lines
 \excludecomment{align*} % Remove align*
 \excludecomment{align} % Remove align
 \excludecomment{equation*} % Remove equation* (does not work)
 \excludecomment{equation} % Remove equation
 \excludecomment{eqnarray*} % Remove eqnarray*
 \excludecomment{eqnarray} % Remove eqnarray
 \excludecomment{acknowledgments} % Remove acknowledgments
 \def\maketitle{} % Remove title and abstract
}{
}

\begin{document}
\ifthenelse{\boolean{wordcount}}{}{
\begin{flushright}
%Belle II preprint 2022-005\\
%KEK preprint 2022-26\\
\end{flushright}
\begin{center}
%(To appear in PRD Letters)
\end{center}}

\title{% WRITE THE TITLE IN THIS FILE
Measurement of the \Omgc lifetime at \belletwo}

\ifthenelse{\boolean{wordcount}}{}
{%%% Paper:    Omega_c lifetime
%%% Journal:  Physical Review D (L)
%%% Contacts: A. Di Canto, N.K. Nisar
%%% ====================================================================
%%% Use \input{pub012-orcid} to insert this material into your latex file.
  \author{F.~Abudin{\'e}n\,\orcidlink{0000-0002-6737-3528}} % 2250
  \author{I.~Adachi\,\orcidlink{0000-0003-2287-0173}} % 2590
% \author{K.~Adamczyk\,\orcidlink{0000-0001-6208-0876}} % 2239
  \author{L.~Aggarwal\,\orcidlink{0000-0002-0909-7537}} % 10083
% \author{P.~Ahlburg\,\orcidlink{0000-0002-9832-7604}} % 2367
  \author{H.~Ahmed\,\orcidlink{0000-0003-3976-7498}} % 11323
% \author{J.~K.~Ahn\,\orcidlink{0000-0002-5795-2243}} % 7423
  \author{H.~Aihara\,\orcidlink{0000-0002-1907-5964}} % 2223
  \author{N.~Akopov\,\orcidlink{0000-0002-4425-2096}} % 9443
  \author{A.~Aloisio\,\orcidlink{0000-0002-3883-6693}} % 2194
% \author{F.~Ameli\,\orcidlink{0000-0001-5435-0450}} % 4683
% \author{L.~Andricek\,\orcidlink{0000-0003-1755-4475}} % 2098
  \author{N.~Anh~Ky\,\orcidlink{0000-0003-0471-197X}} % 2218
  \author{D.~M.~Asner\,\orcidlink{0000-0002-1586-5790}} % 4684
% \author{H.~Atmacan\,\orcidlink{0000-0003-2435-501X}} % 2538
% \author{V.~Aulchenko\,\orcidlink{0000-0002-5394-4406}} % 8183
  \author{T.~Aushev\,\orcidlink{0000-0002-6347-7055}} % 3747
% ...
% 13403
\collaboration{The Belle II Collaboration}
}
%{\input{pub012}}

\begin{abstract}
% WRITE THE ABSTRACT IN THIS FILE
We report on a measurement of the \Omgc lifetime using $\Omgc\to\Omg\pip$ decays reconstructed in $\epem\to\ccbar$ data collected by the \belletwo experiment and corresponding to \lumi of integrated luminosity. The result, $\tau(\Omgc)=\tauOmgc\pm\tauOmgcStat\stat\pm\tauOmgcSyst\syst\fs$, agrees with recent measurements indicating that the \Omgc is not the shortest-lived weakly decaying charmed baryon.
\end{abstract}

\maketitle

The lifetime hierarchy of beauty hadrons is accurately predicted using the so-called heavy-quark expansion, which expresses the decay rate of heavy hadrons as an expansion in inverse powers of the heavy-quark mass $m_b$~\cite{Neubert:1997gu,Uraltsev:2000qw,Lenz:2013aua,Lenz:2014jha,Kirk:2017juj,Cheng:2018rkz}. An accurate prediction of the hierarchy of charmed hadrons is more challenging because higher-order terms in $1/m_c$ and contributions from spectator quarks cannot be neglected and result in larger uncertainties. While the lifetimes of charmed mesons are known to high precision, the lifetimes of charmed baryons are less well measured~\cite{pdg}.

Since its lifetime was first measured in 1995~\cite{E687:1995cvt,WA89:1995lbz}, the \Omgc baryon was believed to be the shortest lived among the four singly charmed baryons that decay weakly~\cite{pdg2018}, in agreement with theoretical expectations~\cite{Shifman:1986mx,Guberina:1986gd}. In 2018, using $\Omgc\to pK^-K^-\pi^+$ decays originating from semileptonic $b$-hadron decays, the LHCb collaboration measured the \Omgc lifetime to be $268\pm24\pm10\pm2\fs$, where the uncertainties are statistical, systematic, and from the \Dp lifetime used as normalization~\cite{Aaij:2018dso}. This value is nearly four times larger than, and inconsistent with, the previous world average of $69\pm12\fs$~\cite{pdg2018}, resulting in the new lifetime hierarchy $\tau(\Xi_c^0) < \tau(\Lambda_c^+) < \tau(\Omgc) < \tau(\Xi_c^+)$. Another recent measurement from LHCb using promptly produced $\Omgc\to pK^-K^-\pi^+$ decays confirms their previous result with better precision, $276.5\pm13.4\pm4.4\pm0.7\fs$, where the last uncertainty is from the \Dz lifetime used as normalization~\cite{LHCb:2021vll}. No independent experimental confirmation of the LHCb results exists. Why the heavy-quark expansion failed to predict the newly observed hierarchy has been debated~\cite{Cheng:2021vca}. However, recently an updated calculation shows that the heavy-quark expansion can satisfactorily describe the measured lifetimes~\cite{Gratrex:2022xpm}.

In this Letter, we report on a measurement of the \Omgc lifetime using $\mbox{\Omgc\to\Omg\pip}$ decays reconstructed in $\epem\to\ccbar$ events at \belletwo. Charge-conjugated decays are included throughout this Letter. The \epem collision data used are collected at center-of-mass energies at or near the $\Upsilon(4S)$ mass and correspond to an integrated luminosity of \lumi. Assuming a lifetime consistent with the LHCb measurement, \Omgc baryons produced in $\epem\to\ccbar$ events at \belletwo have a Lorentz boost that, on average, displaces their decay vertices by 100\mum from the \epem interaction point (IP), where they are produced. The decay time is measured from the projection of the displacement $\vec{L}$ along the direction of the momentum $\vec{p}$, as $t = m\vec{L}\cdot\vec{p}/|\vec{p}|^2$, where $m$ is the known mass of the \Omgc baryon~\cite{pdg}. The decay-time uncertainty \sigmat is calculated by propagating the uncertainties in $\vec{L}$ and $\vec{p}$, including their correlations. The lifetime is determined using a fit to the $(t,\sigmat)$ distributions of the reconstructed \Omgc candidates. To minimize bias, an arbitrary and unknown lifetime offset is applied to the data. The offset is revealed only after we finalized the entire analysis procedure and determined all uncertainties.

The \belletwo detector~\cite{Abe:2010gxa} is built around the collision point of the SuperKEKB asymmetric-energy \epem collider~\cite{Akai:2018mbz} and consists of subsystems arranged in a cylindrical geometry around the beam pipe. The innermost is a tracking subsystem consisting of a two-layer silicon-pixel detector (PXD) surrounded by a four-layer double-sided silicon-strip detector (SVD) and a 56-layer central drift chamber (CDC). Only 15\% of the azimuthal angle is covered by the second PXD layer for the collection of these data. A time-of-propagation counter in the barrel and an aerogel ring-imaging Cherenkov detector in the forward end cap provide information used for the identification of charged particles. An electromagnetic calorimeter consisting of CsI(Tl) crystals fills the remaining volume inside a $1.5\,\rm{T}$ superconducting solenoid and provides energy and timing measurements for photons and electrons. A \KL and muon detection subsystem is installed in the iron flux return of the solenoid. The $z$ axis of the laboratory frame is defined as the central axis of the solenoid, with its positive direction defined as the direction opposite the positron beam.

Events are reconstructed using the Belle II software framework~\cite{Kuhr:2018lps,basf2-zenodo} using selection requirements that ensure large signal efficiency and avoid biases on decay time or variation of the signal efficiency as a function of decay time, as verified in simulation. The simulation uses \textsc{KKMC}~\cite{Jadach:1999vf} to generate quark-antiquark pairs from \epem collisions, \textsc{PYTHIA8}~\cite{Sjostrand:2014zea} to simulate the quark hadronization, \textsc{EVTGEN}~\cite{Lange:2001uf} to decay the hadrons, and \textsc{GEANT4}~\cite{Agostinelli:2002hh} to simulate the detector response. 

\begin{figure}[t!]
\centering
\includegraphics[width=0.5\textwidth]{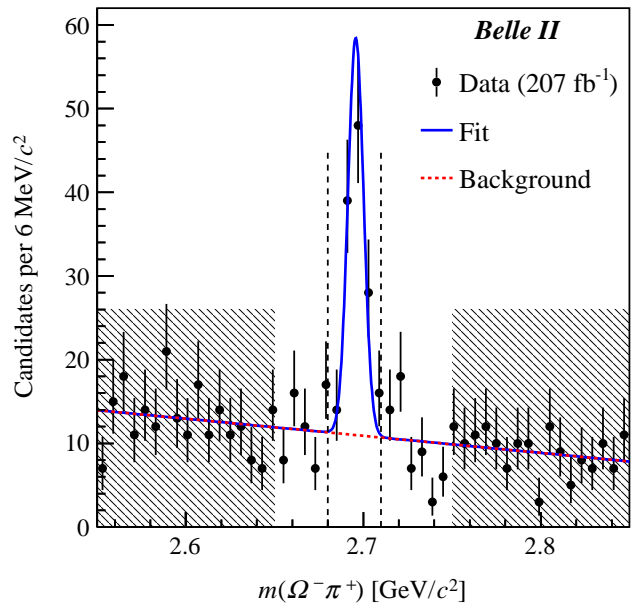}\\
\caption{Mass distribution for \OmgcToOmgPi candidates with fit projections overlaid. The vertical dashed lines enclose the signal region; the shaded area indicates the sideband.\label{fig:massfit}}
\end{figure}

Events enriched in signal \OmgcToOmgPi decays, with $\Omg\to\Lb(\to p\pim) K^-$, are selected by rejecting events consistent with Bhabha scattering and by requiring at least three charged particles, with transverse momenta greater than 200\mevc, that are consistent with originating from the \epem interaction. These charged particles are not required to belong to the \OmgcToOmgPi decay. Candidate $\Lb\to p\pim$ decays are formed using pairs of oppositely charged particles, one of which must be identified as a proton. The decay vertex of the \Lb candidate is required to be more than 0.35\cm away from the IP. The \Lb candidates are combined with negatively charged kaon candidates having transverse momenta greater than 0.15\gevc to form $\Omg\to\Lb K^-$ decays. The \Omg decay vertex must lie between the \Lb vertex and the IP and be at least 0.5\mm from the IP. For both the \Lb and the \Omg candidates, the angle between its momentum and its displacement from the IP must be smaller than $90\degrees$. Candidate \OmgcToOmgPi decays are formed by combining the selected \Omg candidates with positively charged particles that are consistent with originating from the \epem interaction and have momenta greater than $0.5\gevc$. We require the scaled momentum of the \Omgc candidate be larger than $0.6$. The scaled momentum is $\pcms/\sqrt{s/4-\mOmgc^2}$, where \pcms is the momentum of the \Omgc candidate in the \epem center-of-mass system, $s$ is the squared center-of-mass energy, and \mOmgc is the reconstructed \Omgc mass. The scaled momentum requirement eliminates \Omgc candidates originating from decays of \B mesons and greatly suppresses combinatorial background. A decay-chain vertex fit constrains the tracks according to the decay topology and constrains the \Omgc candidate to originate from the \epem interaction region~\cite{treefitter}. The interaction region has typical dimensions of 250\mum along the $z$ axis and of 10\mum and 0.3\mum in the two directions transverse to the $z$ axis. Its position and size vary over time and are measured using $\epem\to\mup\mun$ events. Only candidates with fit probabilities larger than $0.001$ and with \sigmat values smaller than $1.0\ps$ are retained for further analysis. The vertex fit updates the track parameters of the final-state particles, and the updated parameters are used in the subsequent analysis. The \Lb and \Omg candidates are required to have masses within approximately three units of mass resolution (or standard deviations) of their known values~\cite{pdg}. The mass of the \Omgc candidate must be in the range $[2.55,2.85]\gevcc$. After these requirements, about 0.5\% of events have multiple \Omgc candidates; for these events, the candidate with the highest vertex-fit probability is retained. An unbinned maximum likelihood fit to the \mOmgc distribution is used to determine the signal purity in the signal region defined by $2.68<\mOmgc<2.71\gevcc$ (Fig.~\ref{fig:massfit}). In the fit, the \Omgc signal is modeled with a Gaussian distribution, and the background is modeled with a straight line. The signal region contains approximately 132 candidates with a signal purity of $(66.5\pm3.3)\%$.

The lifetime is determined using a maximum-likelihood fit to the unbinned $(t,\sigmat)$ distribution of the candidates populating the signal region. The likelihood is defined as
\begin{multline*}
L(f_{\rm s},\theta)=G(f_{\rm s}|0.665,0.033)\\
\prod_i \left[f_{\rm s}P_{\rm s}(t_i,\sigma_{t\,i}|\theta) + (1-f_{\rm s}) P_{\rm b}(t_i,\sigma_{t\,i}|\theta)\right],
\end{multline*}
where $i$ runs over the candidates and $\theta$ is a short-hand notation for the set of fit parameters, which are specified in the following. The signal fraction $f_{\rm s}$ is constrained to the value measured in the \mOmgc fit with the Gaussian distribution $G(f_{\rm s}|0.665,0.033)$. The signal probability density function (PDF) is the convolution of an exponential distribution in $t$ with a Gaussian resolution function that depends on \sigmat, multiplied by the PDF of \sigmat,
\begin{multline*}
P_{\rm s}(t,\sigmat|\tau,b,s) = P_{\rm s}(t|\sigmat,\tau,b,s)\,P_{\rm s}(\sigmat) \\
\propto \int_0^\infty e^{-t'/\tau} G(t-t'|b,s\sigmat) dt'\,P_{\rm s}(\sigmat)\,.
\end{multline*}
The resolution function's mean $b$ is a free parameter of the fit to account for a possible bias in the determination of the decay time; its width is the per-candidate \sigmat scaled by a free parameter $s$ to account for a possible misestimation of the decay-time uncertainty. The background in the signal region is empirically modeled from data with \mOmgc in the \emph{sideband} $[2.55,2.65]\cup[2.75,2.85]\gevcc$ (Fig.~\ref{fig:massfit}). The sideband is assumed to contain exclusively background candidates and be representative of the background in the signal region, as verified in simulation. The background PDF is the conditional PDF of $t$ given \sigmat multiplied by the PDF of \sigmat, $P_{\rm b}(t,\sigmat|\theta) = P_{\rm b}(t|\sigmat,\theta)\,P_{\rm b}(\sigmat)$. The distribution in $t$ is the sum of a $\delta$ function at zero and an exponential component with lifetime $\tau_{\rm b}$, both convolved with a Gaussian resolution function having a free mean $b_{\rm b}$ and a width corresponding to \sigmat scaled by a free parameter $s_{\rm b}$,
\begin{multline*}
P_{\rm b}(t|\sigmat,\tau_b,f_{\tau_{\rm b}},b_{\rm b},s_{\rm b})
= (1-f_{\tau_b})G(t|b_{\rm b},s_{\rm b}\sigmat) \\
+ f_{\tau_{\rm b}}P_{\rm b}(t|\sigmat,\tau_{\rm b},b_{\rm b},s_{\rm b})\,,
\end{multline*}
where $f_{\tau_{\rm b}}$ is the fraction of the exponential component relative to the total background and
\begin{multline*}
P_{\rm b}(t|\sigmat,\tau_{\rm b},b_{\rm b},s_{\rm b})
\propto \int_0^\infty e^{-t'/\tau_{\rm b}} G(t-t'|b_{\rm b},s_{\rm b}\sigmat) dt'\,.
\end{multline*}
To better constrain the background parameters, a simultaneous fit to the candidates in the signal region and the sideband is performed. The PDFs of \sigmat, which differ between signal and background, are histogram templates derived directly from the data. The signal template is derived from the candidates in the signal region after subtracting the scaled distribution of the sideband data. The background template is obtained directly from the sideband data. No direct input from simulation is used in the fit.

\begin{figure}[ht!]
\centering
\includegraphics[width=0.5\textwidth]{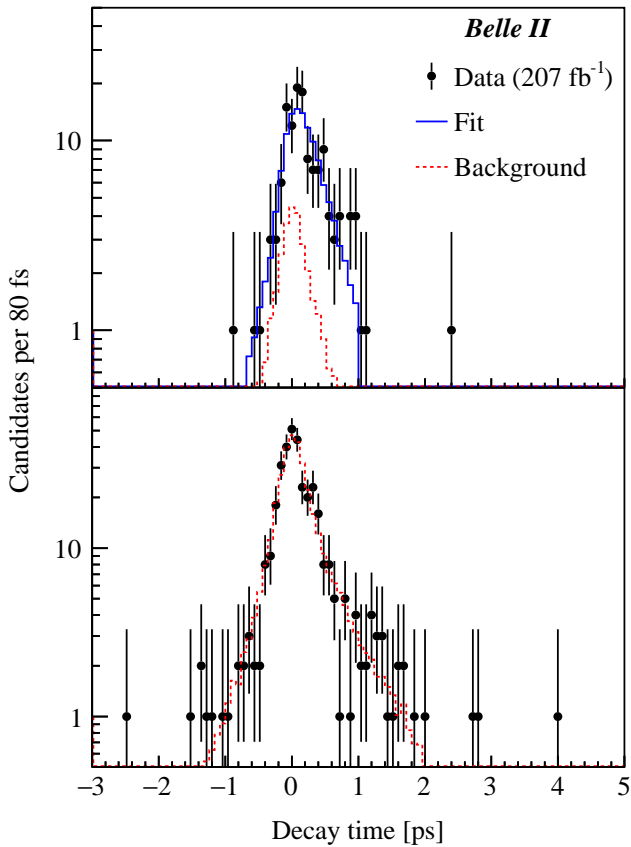}\\
\caption{Decay-time distributions for \OmgcToOmgPi candidates populating (top) the signal region and (bottom) the sideband with fit projections overlaid.\label{fig:lifetime-fit-t}}
\end{figure}

\begin{figure}[ht!]
\centering
\includegraphics[width=0.5\textwidth]{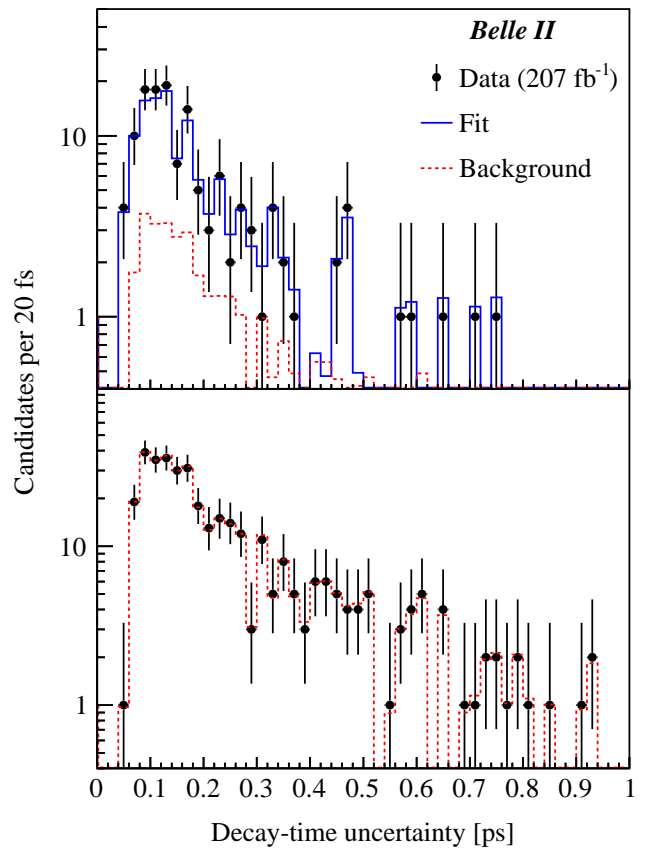}\\
\caption{Decay-time-uncertainty distributions for \OmgcToOmgPi candidates populating (top) the signal region and (bottom) the sideband with fit projections overlaid.\label{fig:lifetime-fit-sigmat}}
\end{figure}

The distributions of decay time and decay-time uncertainty are shown in Figs.~\ref{fig:lifetime-fit-t} and \ref{fig:lifetime-fit-sigmat} with fit projections overlaid. The \Omgc lifetime is measured to be $\tauOmgc\pm\tauOmgcStat\fs$, the mean of the signal resolution function is $b=-18\pm41\fs$, and the scaling factor of the width is $s=1.35\pm0.20$, where the uncertainties are statistical only.

\begin{table}[t]
\centering
\caption{Systematic uncertainties.\label{tab:syst}}
\begin{tabular}{lc}
\hline\hline
Source & Uncertainty (fs) \\
\hline
Fit bias           &  $\phantom{0}3.4$ \\
Resolution model   &  $\phantom{0}6.2$ \\
Background model   &  $\phantom{0}8.3$ \\
Detector alignment &  $\phantom{0}1.6$ \\
Momentum scale     &  $\phantom{0}0.2$ \\
Input \Omgc mass   &  $\phantom{0}0.2$ \\
\hline
Total & 11.0  \\
\hline\hline
\end{tabular}
\end{table}

The following sources of systematic uncertainties are considered: fit bias, resolution model, treatment of background contamination, imperfect alignment of the tracking detectors, and uncertainties in the momentum scale and in the input \Omgc mass. Table~\ref{tab:syst} lists all contributions and their total, calculated as the sum in quadrature of the individual contributions.

The lifetime fit is tested on data generated by randomly sampling the fit PDF with parameters fixed to the values found in the fit to the data and with lifetime values varied between 60\fs and 300\fs. One thousand pseudoexperiments, each the same size as the data, are generated for each tested lifetime value. A $-3.4\fs$ bias is observed for lifetime values close to the fit result of $\tauOmgc\fs$. The bias is mostly due to the small sample size and reduces when simulating larger sizes. Its absolute value is assigned as a symmetric systematic uncertainty. 

Simulation shows that the resolution function has tails that are inconsistent with a Gaussian model. The effect on the measured lifetime due to using our imperfect resolution model is quantified using one thousand samples of signal-only simulated decays, each the same size as the data. The samples are obtained by resampling, with replacement, from a sample of simulated \epem collisions corresponding to five times the data size. For each sample the fit is performed and the measured lifetime is compared to the true lifetime of the parent simulation sample. The average difference between measured and true lifetimes, $2.8\fs$, is corrected for the known fit bias of $-3.4\fs$ and the resulting value, 6.2\fs, is assigned as a systematic uncertainty due to the imperfect resolution model.

For signal decays, the decay-time resolution function has a mean that depends nearly linearly on the candidate mass, and is expected to average out for a symmetric range of candidate masses. We check that the associated uncertainty in the measured lifetime is negligible by varying the boundaries of the signal region.

In simulation, the $(t,\sigmat)$ distribution of the candidates in the sideband describes the background candidates in the signal region well. The same might not hold for the data and this could bias the result. To quantify this bias, we generate and fit to one thousand pseudoexperiments, each the same size and with the same signal-to-background proportion as that of the data. In the generation, signal and background candidates populating the signal region are sampled from the fit PDFs, using input parameters equal to those determined from the fit to the data. Generated background candidates in the signal region thus feature the same $(t,\sigmat)$ distribution as the data. In contrast, candidates in the sideband are sampled from simulated \epem collisions. In this manner, the pseudoexperiments feature sideband data that differ from the background in the signal region with the same level of disagreement as observed between data and simulation. The averaged difference between the measured and generated lifetimes, corrected for the previously estimated biases due to the fit and to the resolution model, is $6.2\pm1.9\fs$. Various definitions of the sideband are tried: $[2.55, 2.64]\cup[2.76,2.85]\gevcc$, $[2.55, 2.66]\cup[2.74,2.85]\gevcc$, $[2.55, 2.65]\gevcc$, and $[2.75,2.85]\gevcc$. The latter region shows a significant deviation in fitted lifetime from the nominal result. The deviation, 8.3\fs, is consistent with the pseudoexperiments study and is assigned as a systematic uncertainty due to the modeling of the background $(t,\sigmat)$ distribution.

In the lifetime fit, the fraction of background candidates in the signal region is constrained by the result of the fit to the \mOmgc distribution. When we change this background fraction to values obtained from fitting to the \mOmgc distribution with alternative signal and background PDFs, the change in the measured lifetime is negligible.

In Belle II, track parameters are periodically calibrated to correct for misalignment and deformation of internal components of the PXD and SVD, and for the relative alignments of the PXD, SVD, and CDC. Misalignment can bias the measurement of the decay lengths and hence of the decay times. To quantify the effect of possible residual misalignment on the measured lifetime, large samples of signal decays are simulated with various misalignment configurations. Lifetime residuals with respect to perfectly aligned simulation are estimated, and their root mean square, $1.6\fs$, is assigned as a systematic uncertainty due to possible detector misalignment. 

Uncertainties in the knowledge of the absolute momentum scale and in the world-average value of the \Omgc mass~\cite{pdg} each result in a $0.2\fs$ uncertainty in the lifetime.

Consistency of the results is tested by repeating the full analysis in subsets of the data split according to data-taking periods and conditions, \Omgc momentum and flight direction, charm flavor, and \Omg flight length. In all cases, the variations of the results are consistent with statistical fluctuations. To check that the best-candidate selection in events with multiple candidates does not affect the result, the measurement is repeated with randomly selecting a single candidate, removing all events with multiple candidates, or keeping all candidates. No significant variation in the measured lifetime is observed. The measurement is also repeated with the fit range varied to exclude candidates in the tails of the $(t,\sigmat)$ distribution, with no significant deviation in the resulting lifetime from the nominal result.

In conclusion, we report on a measurement of the \Omgc lifetime using $\epem\to\ccbar$ data collected by the \belletwo experiment corresponding to an integrated luminosity of \lumi. This measurement, 
$$\tau(\Omgc)=\tauOmgc\pm\tauOmgcStat\stat\pm\tauOmgcSyst\syst\fs\,,$$
is consistent with the LHCb average of \mbox{$274.5\pm12.4\fs$}~\cite{LHCb:2021vll}, and inconsistent at 3.4 standard deviations with the pre-LHCb world average of $69\pm12\fs$~\cite{pdg2018}. The Belle II result, therefore, confirms that the \Omgc is not the shortest-lived weakly decaying charmed baryon.

\begin{acknowledgments}
% Policy from October 20, 2022
This work, based on data collected using the Belle II detector, which was built and commissioned prior to March 2019, was supported by
%Armenia
Science Committee of the Republic of Armenia Grant No.~20TTCG-1C010;
%Australia
Australian Research Council and research Grants
No.~DE220100462,
No.~DP180102629,
No.~DP170102389,
No.~DP170102204,
No.~DP150103061,
No.~FT130100303,
No.~FT130100018,
and
No.~FT120100745;
%Austria
Austrian Federal Ministry of Education, Science and Research,
Austrian Science Fund
No.~P~31361-N36
and
No.~J4625-N,
and
Horizon 2020 ERC Starting Grant No.~947006 ``InterLeptons'';
%Canada
Natural Sciences and Engineering Research Council of Canada, Compute Canada and CANARIE;
%China
Chinese Academy of Sciences and research Grant No.~QYZDJ-SSW-SLH011,
National Natural Science Foundation of China and research Grants
No.~11521505,
No.~11575017,
No.~11675166,
No.~11761141009,
No.~11705209,
and
No.~11975076,
LiaoNing Revitalization Talents Program under Contract No.~XLYC1807135,
Shanghai Pujiang Program under Grant No.~18PJ1401000,
and the CAS Center for Excellence in Particle Physics (CCEPP);
%Czech Republic
the Ministry of Education, Youth, and Sports of the Czech Republic under Contract No.~LTT17020 and
Charles University Grant No.~SVV 260448 and
the Czech Science Foundation Grant No.~22-18469S;
%EU
European Research Council, Seventh Framework PIEF-GA-2013-622527,
Horizon 2020 ERC-Advanced Grants No.~267104 and No.~884719,
Horizon 2020 ERC-Consolidator Grant No.~819127,
Horizon 2020 Marie Sklodowska-Curie Grant Agreement No.~700525 "NIOBE"
and
No.~101026516,
and
Horizon 2020 Marie Sklodowska-Curie RISE project JENNIFER2 Grant Agreement No.~822070 (European grants);
%France
L'Institut National de Physique Nucl\'{e}aire et de Physique des Particules (IN2P3) du CNRS (France);
%Germany
BMBF, DFG, HGF, MPG, and AvH Foundation (Germany);
%India
Department of Atomic Energy under Project Identification No.~RTI 4002 and Department of Science and Technology (India);
%Israel
Israel Science Foundation Grant No.~2476/17,
U.S.-Israel Binational Science Foundation Grant No.~2016113, and
Israel Ministry of Science Grant No.~3-16543;
%Italy
Istituto Nazionale di Fisica Nucleare and the research grants BELLE2;
%Japan
Japan Society for the Promotion of Science, Grant-in-Aid for Scientific Research Grants
No.~16H03968,
No.~16H03993,
No.~16H06492,
No.~16K05323,
No.~17H01133,
No.~17H05405,
No.~18K03621,
No.~18H03710,
No.~18H05226,
No.~19H00682, % Niigata
No.~22H00144,
No.~26220706,
and
No.~26400255,
the National Institute of Informatics, and Science Information NETwork 5 (SINET5), 
and
the Ministry of Education, Culture, Sports, Science, and Technology (MEXT) of Japan;  
%Korea
National Research Foundation (NRF) of Korea Grants
No.~2016R1\-D1A1B\-02012900,
No.~2018R1\-A2B\-3003643,
No.~2018R1\-A6A1A\-06024970,
No.~2018R1\-D1A1B\-07047294,
No.~2019R1\-I1A3A\-01058933,
No.~2022R1\-A2C\-1003993,
and
No.~RS-2022-00197659,
Radiation Science Research Institute,
Foreign Large-size Research Facility Application Supporting project,
the Global Science Experimental Data Hub Center of the Korea Institute of Science and Technology Information
and
KREONET/GLORIAD;
%Malaysia
Universiti Malaya RU grant, Akademi Sains Malaysia, and Ministry of Education Malaysia;
%Mexico
% CINVESTAV-IPN, UNAM, UAS, BUAP and CONACYT are funded under
Frontiers of Science Program Contracts
No.~FOINS-296,
No.~CB-221329,
No.~CB-236394,
No.~CB-254409,
and
No.~CB-180023, and No.~SEP-CINVESTAV research Grant No.~237 (Mexico);
%Poland
the Polish Ministry of Science and Higher Education and the National Science Center;
%Russia
the Ministry of Science and Higher Education of the Russian Federation,
Agreement No.~14.W03.31.0026, and
the HSE University Basic Research Program, Moscow;
%Saudi Arabia
University of Tabuk research Grants
No.~S-0256-1438 and No.~S-0280-1439 (Saudi Arabia);
%Slovenia
Slovenian Research Agency and research Grants
No.~J1-9124
and
No.~P1-0135;
%Spain
Agencia Estatal de Investigacion, Spain
Grant No.~RYC2020-029875-I
and
Generalitat Valenciana, Spain
Grant No.~CIDEGENT/2018/020
%Taiwan
Ministry of Science and Technology and research Grants
No.~MOST106-2112-M-002-005-MY3
and
No.~MOST107-2119-M-002-035-MY3,
and the Ministry of Education (Taiwan);
%Thailand
Thailand Center of Excellence in Physics;
%Turkey
TUBITAK ULAKBIM (Turkey);
%Ukraine
National Research Foundation of Ukraine, project No.~2020.02/0257,
and
Ministry of Education and Science of Ukraine;
%USA
the U.S. National Science Foundation and research Grants
No.~PHY-1913789 % Indiana CEEM
and
No.~PHY-2111604, % Luther
and the U.S. Department of Energy and research Awards
No.~DE-AC06-76RLO1830, % PNNL
No.~DE-SC0007983, % Wayne State
No.~DE-SC0009824, % Florida
No.~DE-SC0009973, % VPI
No.~DE-SC0010007, % Duke
No.~DE-SC0010073, % South Carolina
No.~DE-SC0010118, % Carnegie Mellon
No.~DE-SC0010504, % Hawaii
No.~DE-SC0011784, % Cincinnati
No.~DE-SC0012704, % BNL
No.~DE-SC0019230, % Duke
No.~DE-SC0021274, % Mississippi
No.~DE-SC0022350; % Louisville
%last group
and
%Vietnam
the Vietnam Academy of Science and Technology (VAST) under Grant No.~DL0000.05/21-23.

% Policy from October 20, 2022
These acknowledgements are not to be interpreted as an endorsement of any statement made
by any of our institutes, funding agencies, governments, or their representatives.

We thank the SuperKEKB team for delivering high-luminosity collisions;
the KEK cryogenics group for the efficient operation of the detector solenoid magnet;
the KEK computer group and the NII for on-site computing support and SINET6 network support;
and the raw-data centers at BNL, DESY, GridKa, IN2P3, INFN, and the University of Victoria for offsite computing support.

\end{acknowledgments}

% Produces the bibliography via BibTeX.
\ifthenelse{\boolean{wordcount}}%
{ \bibliographystyle{unsrt}
  \nobibliography{references}}
{ \bibliographystyle{belle2-note}
\bibliography{references}}

\providecommand{\href}[2]{#2}\begingroup\raggedright\begin{thebibliography}{10}

\bibitem{Neubert:1997gu}
M.~Neubert, {\em {$B$ decays and the heavy quark expansion}\/},
  \href{http://dx.doi.org/10.1142/9789812812667_0003}{Adv. Ser. Direct. High
  Energy Phys. {\bf 15} (1998)  239},
  \href{http://arxiv.org/abs/hep-ph/9702375}{{\tt arXiv:hep-ph/9702375}}.

\bibitem{Uraltsev:2000qw}
N.~Uraltsev, \href{http://dx.doi.org/10.1142/9789812810458_0034}{{\em {Topics
  in the heavy quark expansion}\/}, } in {\em At the frontier of Particle
  Physics}, M.~Shifman and B.~Ioffe, eds.
\newblock 2001.
\newblock \href{http://arxiv.org/abs/hep-ph/0010328}{{\tt
  arXiv:hep-ph/0010328}}.

\bibitem{Lenz:2013aua}
A.~Lenz and T.~Rauh, {\em {$D$-meson lifetimes within the heavy quark
  expansion}\/},  \href{http://dx.doi.org/10.1103/PhysRevD.88.034004}{Phys.
  Rev. D {\bf 88} (2013)  034004}, \href{http://arxiv.org/abs/1305.3588}{{\tt
  arXiv:1305.3588 [hep-ph]}}.

\bibitem{Lenz:2014jha}
A.~Lenz, {\em {Lifetimes and heavy quark expansion}\/},
  \href{http://dx.doi.org/10.1142/S0217751X15430058}{Int. J. Mod. Phys. A {\bf
  30} (2015)  1543005}, \href{http://arxiv.org/abs/1405.3601}{{\tt
  arXiv:1405.3601 [hep-ph]}}.

\bibitem{Kirk:2017juj}
M.~Kirk, A.~Lenz, and T.~Rauh, {\em {Dimension-six matrix elements for meson
  mixing and lifetimes from sum rules}\/},
  \href{http://dx.doi.org/10.1007/JHEP12(2017)068}{JHEP {\bf 12} (2017)  068},
  \href{http://arxiv.org/abs/1711.02100}{{\tt arXiv:1711.02100 [hep-ph]}}.
  Erratum \href{https://doi.org/10.1007/JHEP06(2020)162}{JHEP {\bf 06} (2020)
  162}.

\bibitem{Cheng:2018rkz}
H.-Y. Cheng, {\em {Phenomenological study of heavy hadron lifetimes}\/},
  \href{http://dx.doi.org/10.1007/JHEP11(2018)014}{JHEP {\bf 11} (2018)  014},
  \href{http://arxiv.org/abs/1807.00916}{{\tt arXiv:1807.00916 [hep-ph]}}.

\bibitem{pdg}
P.~A. Zyla et al., {Particle Data Group}, {\em {Review of Particle Physics}\/},
\href{http://dx.doi.org/10.1093/ptep/ptaa104}{PTEP {\bf 2020} (2020)  083C01}.
%%CITATION = INSPIRE-1812251;%%.

\bibitem{E687:1995cvt}
P.~L. Frabetti et al., {E687 collaboration}, {\em {First measurement of the
  lifetime of the \Omgc}\/},
  \href{http://dx.doi.org/10.1016/0370-2693(95)00941-D}{Phys. Lett. B {\bf 357}
  (1995)  678}.

\bibitem{WA89:1995lbz}
M.~I. Adamivich et al., {WA89 collaboration}, {\em {Measurement of the \Omgc
  lifetime}\/},  \href{http://dx.doi.org/10.1016/0370-2693(95)00979-U}{Phys.
  Lett. B {\bf 358} (1995)  151},
  \href{http://arxiv.org/abs/hep-ex/9507004}{{\tt arXiv:hep-ex/9507004}}.

\bibitem{pdg2018}
M.~Tanabashi et al., {Particle Data Group}, {\em Review of Particle Physics\/},
   \href{http://dx.doi.org/10.1103/PhysRevD.98.030001}{Phys. Rev. D {\bf 98}
  (2018)  030001}.

\bibitem{Shifman:1986mx}
M.~A. Shifman and M.~B. Voloshin, {\em {Hierarchy of Lifetimes of Charmed and
  Beautiful Hadrons}\/},  Sov. Phys. JETP {\bf 64} (1986)  698.

\bibitem{Guberina:1986gd}
B.~Guberina, R.~Ruckl, and J.~Trampetic, {\em {Charmed Baryon Lifetime
  Differences}\/},  \href{http://dx.doi.org/10.1007/BF01411150}{Z. Phys. C {\bf
  33} (1986)  297}.

\bibitem{Aaij:2018dso}
R.~Aaij et al., {LHCb collaboration}, {\em {Measurement of the $\Omega_c^0$
  baryon lifetime}\/},
  \href{http://dx.doi.org/10.1103/PhysRevLett.121.092003}{Phys. Rev. Lett. {\bf
  121} (2018)  092003}, \href{http://arxiv.org/abs/1807.02024}{{\tt
  arXiv:1807.02024 [hep-ex]}}.

\bibitem{LHCb:2021vll}
R.~Aaij et al., {LHCb collaboration}, {\em {Measurement of the lifetimes of
  promptly produced \Omgc and $\Xi_c^0$ baryons}\/},
  \href{http://dx.doi.org/10.1016/j.scib.2021.11.022}{Sci. Bull. {\bf 67}
  (2022) no.~5, 479--487}, \href{http://arxiv.org/abs/2109.01334}{{\tt
  arXiv:2109.01334 [hep-ex]}}.

\bibitem{Cheng:2021vca}
H.-Y. Cheng, {\em {The strangest lifetime: A bizarre story of
  $\tau(\Omega_c^0)$}\/},
  \href{http://dx.doi.org/10.1016/j.scib.2021.11.025}{Sci. Bull. {\bf 67}
  (2022) no.~5, 445--447}, \href{http://arxiv.org/abs/2111.09566}{{\tt
  arXiv:2111.09566 [hep-ph]}}.

\bibitem{Gratrex:2022xpm}
J.~Gratrex, B.~Meli\'c, and I.~Ni\v{s}and\v{z}i\'c, {\em {Lifetimes of singly
  charmed hadrons}\/},  \href{http://arxiv.org/abs/2204.11935}{{\tt
  arXiv:2204.11935 [hep-ph]}}.

\bibitem{Abe:2010gxa}
T.~Abe et al., {Belle II collaboration}, {\em {Belle II Technical Design
  Report}\/},
\href{http://arxiv.org/abs/1011.0352}{{\tt arXiv:1011.0352 [physics.ins-det]}}.
%%CITATION = ARXIV:1011.0352;%%.

\bibitem{Akai:2018mbz}
K.~Akai, K.~Furukawa, and H.~Koiso, {\em {SuperKEKB Collider}\/},
  \href{http://dx.doi.org/10.1016/j.nima.2018.08.017}{Nucl. Instrum. Meth. {\bf
  A907} (2018)  188},
\href{http://arxiv.org/abs/1809.01958}{{\tt arXiv:1809.01958
  [physics.acc-ph]}}.
%%CITATION = ARXIV:1809.01958;%%.

\bibitem{Kuhr:2018lps}
T.~Kuhr, C.~Pulvermacher, M.~Ritter, T.~Hauth, and N.~Braun, {Belle II
  framework software group}, {\em {The Belle II Core Software}\/},
  \href{http://dx.doi.org/10.1007/s41781-018-0017-9}{Comput. Softw. Big Sci.
  {\bf 3} (2019)  1},
\href{http://arxiv.org/abs/1809.04299}{{\tt arXiv:1809.04299
  [physics.comp-ph]}}.
%%CITATION = ARXIV:1809.04299;%%.

\bibitem{basf2-zenodo}
{The Belle II Collaboration}, {\em {Belle II Analysis Software Framework
  (basf2)}\/},  \url{https://doi.org/10.5281/zenodo.5574115}.

\bibitem{Jadach:1999vf}
S.~Jadach, B.~F.~L. Ward, and Z.~W\c{a}s, {\em {The precision Monte Carlo event
  generator KK for two-fermion final states in $e^+e^-$ collisions}\/},
  \href{http://dx.doi.org/10.1016/S0010-4655(00)00048-5}{Comput. Phys. Commun.
  {\bf 130} (2000)  260},
\href{http://arxiv.org/abs/hep-ph/9912214}{{\tt arXiv:hep-ph/9912214
  [hep-ph]}}.
%%CITATION = HEP-PH/9912214;%%.

\bibitem{Sjostrand:2014zea}
T.~Sjöstrand, S.~Ask, J.~R. Christiansen, R.~Corke, N.~Desai, P.~Ilten,
  S.~Mrenna, S.~Prestel, C.~O. Rasmussen, and P.~Z. Skands, {\em {An
  Introduction to PYTHIA 8.2}\/},
  \href{http://dx.doi.org/10.1016/j.cpc.2015.01.024}{Comput. Phys. Commun. {\bf
  191} (2015)  159},
\href{http://arxiv.org/abs/1410.3012}{{\tt arXiv:1410.3012 [hep-ph]}}.
%%CITATION = ARXIV:1410.3012;%%.

\bibitem{Lange:2001uf}
D.~J. Lange, {\em {The EvtGen particle decay simulation package}\/},
\href{http://dx.doi.org/10.1016/S0168-9002(01)00089-4}{Nucl. Instrum. Meth.
  {\bf A462} (2001)  152}.
%%CITATION = NUIMA,A462,152;%%.

\bibitem{Agostinelli:2002hh}
S.~Agostinelli et al., {GEANT4 collaboration}, {\em {GEANT4: A Simulation
  toolkit}\/},
\href{http://dx.doi.org/10.1016/S0168-9002(03)01368-8}{Nucl.Instrum.Meth. {\bf
  A506} (2003)  250}.
%%CITATION = NUIMA,A506,250;%%.

\bibitem{treefitter}
J.-F. Krohn et al., {Belle II analysis software group}, {\em {Global decay
  chain vertex fitting at Belle II}\/},
  \href{http://dx.doi.org/10.1016/j.nima.2020.164269}{Nucl. Instrum. Meth. {\bf
  A976} (2020)  164269}, \href{http://arxiv.org/abs/1901.11198}{{\tt
  arXiv:1901.11198 [hep-ex]}}.

\end{thebibliography}\endgroup

\end{document}